\documentclass[%
 aip,
 amsmath,amssymb,
 reprint,%
]{revtex4-1}

\usepackage{graphicx}
\graphicspath{ {./Figures/} }
\usepackage{dcolumn}
\usepackage{bm}
\usepackage{todonotes}
\usepackage[utf8]{inputenc}
\usepackage[T1]{fontenc}
\usepackage{mathptmx}
\usepackage{etoolbox}
\usepackage{subcaption}
\usepackage{float}
\usepackage{siunitx}
\usepackage{hyperref}
\hypersetup{
    pdftex,
    bookmarks=true,
    bookmarksnumbered=true,
    bookmarksopen=true,
    bookmarksopenlevel=2,
    pdfstartview=FitH,
    linktocpage=true,
    colorlinks=true,
    linkcolor=red!50!black,
    urlcolor=blue,
    citecolor=teal,
    pdfpagelabels=true
}
\makeatletter
\def\@email#1#2{%
 \endgroup
 \patchcmd{\titleblock@produce}
  {\frontmatter@RRAPformat}
  {\frontmatter@RRAPformat{\produce@RRAP{*#1\href{mailto:#2}{#2}}}\frontmatter@RRAPformat}
  {}{}
}%
\makeatother
\begin{document}

\preprint{AIP/123-QED}

\title[The following article has been submitted to JVST A. After it is published, it will be found at \href{https://publishing.aip.org/resources/librarians/products/journals/}{link}]{Challenges and Insights in Growing Epitaxial FeSn Thin Films on
\\GaAs(111) substrate Using Molecular Beam Epitaxy}
\author{P. Chatterjee}
\affiliation{ 
Center for Quantum Spintronics, Department of Physics, Norwegian University of Science and Technology (NTNU), NO-7491 Trondheim, Norway
}%

\author{M. Nord}
\affiliation{ 
Department of Physics, Norwegian University of Science and Technology (NTNU), NO-7491 Trondheim, Norway
}%

\author{J. He}
\affiliation{Department of Materials Science and Engineering, Norwegian University of Science and Technology (NTNU), NO-7491 Trondheim, Norway
}%

\author{D. Meier}
\affiliation{ Center for Quantum Spintronics, Department of Physics, Norwegian University of Science and Technology (NTNU), NO-7491 Trondheim, Norway
}%
\affiliation{Department of Materials Science and Engineering, Norwegian University of Science and Technology (NTNU), NO-7491 Trondheim, Norway
}%

\author{C. Brüne}%
\homepage{Corresponding author email: \underline{christoph.brune@ntnu.no}}
\affiliation{ 
Center for Quantum Spintronics, Department of Physics, Norwegian University of Science and Technology (NTNU), NO-7491 Trondheim, Norway
}%

\begin{abstract}
FeSn is a room-temperature antiferromagnet composed of alternating \(\mathrm{Fe_{3}Sn}\) kagome layers and honeycomb Sn layers. Its distinctive lattice allows the formation of linearly dispersing Dirac bands and topological flat bands in its electronic band structure, positioning FeSn as an ideal candidate for investigating the interplay between magnetism and topology. In this study, we investigate the epitaxial growth of FeSn thin films on GaAs(111) substrates by molecular beam epitaxy. A significant challenge in this growth process is the diffusion of Ga and As from the substrate into the deposited films and the diffusion of Fe into the substrate.  This diffusion complicates the formation of a pure FeSn phase. Through a comprehensive analysis—including reflection high energy electron diffraction, high-resolution X-ray diffraction, scanning electron microscopy, transmission electron microscopy, and vibrating sample magnetometry —we demonstrate that the Sn evaporation temperature plays a critical role in influencing the crystallinity, surface morphology, and magnetic behaviour of the films. Our results show that while it is difficult to grow a single-phase FeSn film on GaAs due to diffusion, optimizing the Sn evaporation temperature can enhance the dominance of the FeSn phase, partially overcoming these challenges. 
\end{abstract}

\maketitle

\section{\label{sec:level1}Introduction}
In the evolving landscape of material science, the study of binary metal magnetic kagome compounds \(\mathrm{ T_{m}X_{n}}\) (where T is a 3d transition metal like Fe, Mn, or Co, and X is commonly Sn, Ir, or Ge) has gained significant momentum due to their potential applications in various advanced technologies \cite{zhang2021recent,chowdhury2023kagome}. Specifically, \(\mathrm{ Fe_{m}Sn_{n}} \) compounds have sparked considerable interest in spintronics, due to their ability to exhibit both ferromagnetic and antiferromagnetic phases depending on the stacking sequence of the kagome layers. Five distinct intermetallic compounds in the Fe-Sn family—\(\mathrm{ Fe_{}Sn_{}} \), \( \mathrm{Fe_{3}Sn_{2}} \), \( \mathrm{Fe_{3}Sn_{}}\), \( \mathrm{Fe_{5}Sn_{3}}\), and \( \mathrm{Fe_{}Sn_{2}} \) —have been reported\cite{jannin1963magnetism} and extensively examined\cite{giefers2006high,trumpy1970mossbauer}. Due to its topological properties, FeSn is a promising candidate for antiferromagnetic spintronics among these compounds. Recent progress in controlling and detecting antiferromagnetic states in topological semimetals has opened up new avenues in antiferromagnetic spintronics \cite{vsmejkal2017route,vsmejkal2018topological,vzelezny2018spin,chen2024emerging,bonbien2021topological,he2022topological,bernevig2022progress}. 

\begin{figure}[htb!]
    \centering
    \includegraphics[width=1\linewidth]{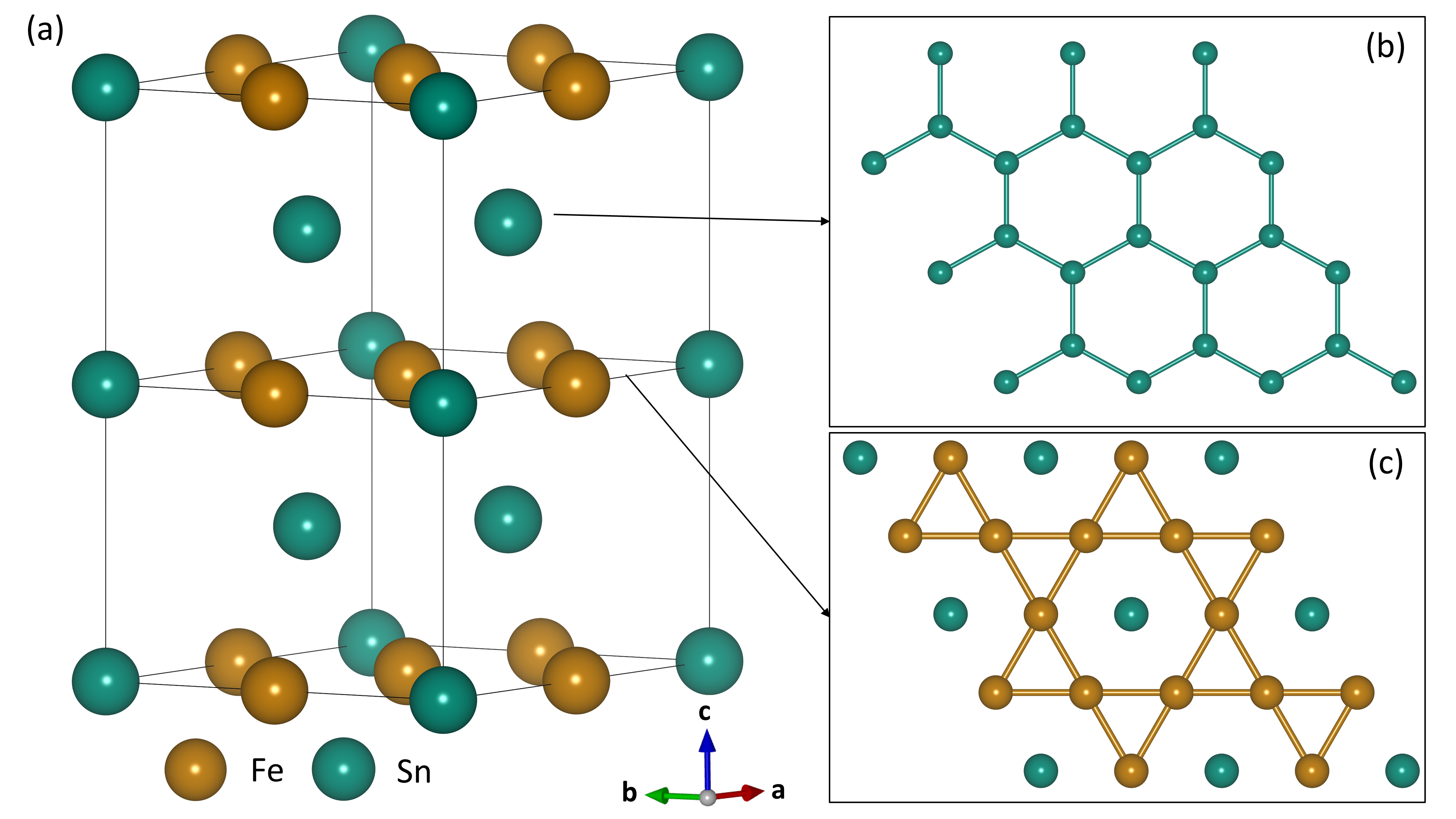}\hfill
    \caption{\textbf{Crystal Structure of FeSn:}\textbf{(a)} Three-dimensional lattice structure of FeSn, showing the arrangement of Fe and Sn atoms. \textbf{(b) }Sn layer which separates the \(\mathrm{Fe_{3}Sn}\) layers.  \textbf{(c)} Top view of the FeSn lattice highlighting the \(\mathrm{Fe_{3}Sn}\) layer, where Fe atoms form the kagome lattice.}
    \label{fig:FeSn}
\end{figure}

\( \mathrm{Fe_{}Sn_{}} \) possesses a hexagonal crystal structure with space group \(\mathrm{P6_3/mmc}\). It is composed of single \(\mathrm{Fe_{3}Sn}\) layers separated by Sn layers, as shown in figure \ref{fig:FeSn}(a) to (c). The lattice constants are a = b = 5.296 Å and c = 4.448 Å \cite{giefers2006high}. Each kagome layer is composed of corner-sharing triangles of Fe atoms with Sn atoms occupying the centre of the hexagon, as shown in figure \ref{fig:FeSn}(c). The stacking order of the spatially decoupled kagome lattice is responsible for the complex magnetic behaviour arising from competing magnetic interactions and geometrical frustration. FeSn exhibits antiferromagnetic ordering with a Néel temperature \(\mathrm{T_{N}\approx 365 K} \)  \cite{giefers2006high}. The ferromagnetic coupling of Fe atoms within the kagome layers and the antiferromagnetic alignment between adjacent kagome layers contribute to its complex magnetic properties, as evidenced in neutron diffraction and Mössbauer experiments \cite{yamaguchi1967neutron,kulshreshtha1981anisotropic,sales2019electronic}. A notable feature of FeSn's band structure is the coexistence of Dirac fermions and flat bands close to the Fermi level \cite{kang2020dirac}. Angle-resolved photoemission spectroscopy (ARPES) measurement on FeSn single crystals confirmed the existence of massless Dirac fermions in bulk and  2D Weyl-like states at the surface \cite{lin2020dirac}. The presence of spin-polarized 2D flat bands at the surface of FeSn thin films has also been confirmed by planar tunnelling spectroscopy and first-principles calculation \cite{han2021evidence}. These unique magnetic and electronic properties of FeSn open up possibilities for application in antiferromagnetic memory devices \cite{tsai2020electrical,tsai2021spin,chen2024emerging,bernevig2022progress}. 

The epitaxial growth of FeSn thin films on suitable substrates is a critical step toward harnessing these properties for technological applications. The GaAs(111) substrate has a lattice mismatch of approximately 6.3\% with FeSn \cite{ohtake2001surface}. GaAs substrates are widely available and relatively inexpensive, making them attractive for epitaxial growth. The intrinsic properties of GaAs, such as its high electron mobility and direct bandgap, make it an attractive candidate for spintronic applications \cite{ohno1998making}. In the context of antiferromagnetic spintronics, GaAs-based heterostructures have attracted significant attention. GaAs is a non-magnetic semiconductor with global inversion asymmetry, which generates spin polarization in the bulk due to the inverse spin galvanic effect (ISGE) \cite{kato2004current,chernyshov2009evidence,silov2004current,wunderlich2005experimental}. This effect can be used to manipulate the antiferromagnetic moments in adjacent layers using electrical currents \cite{jungwirth2016antiferromagnetic,wadley2016electrical,baltz2018antiferromagnetic,gomonay2017concepts}.

The growth of  FeSn thin films by molecular beam epitaxy (MBE) was first reported by H. Inoue \textit{et al.} \cite{inoue2019molecular}. The films were grown on \( \mathrm{SrTiO_{3}}\)(111) (STO) substrate and the X-ray diffraction (XRD) data revealed that the films were oriented in the (001) direction. Epitaxial thin films of FeSn have also been successfully deposited on substrates like Nb-doped STO \cite{han2021evidence}, \(\mathrm{LaAlO_{3}}\)(111) \cite{hong2020molecular}, \(\mathrm{Al_{2}O_{3}}\) (001) \cite{khadka2020high} and Si(111)/(001)\cite{bhattarai2023magnetotransport}. These works lay the groundwork for the present study, where we expand the synthesis towards the growth of FeSn thin films on GaAs(111) substrates.

However, growing FeSn thin films on GaAs(111) is challenging. A persistent challenge in the growth of Fe-based compounds on GaAs substrates is the diffusion of As and Ga into the film and diffusion of Fe into the substrate, which can form secondary intermetallic compounds such as \(\mathrm{FeAs}\), \(\mathrm{Fe_2As}\) and \(\mathrm{Fe_3Ga}\) \cite{schultz2008phase,rahmoune1997analysis}. In this study, we focus on the effect of the Sn evaporation temperature (\(\mathrm{T_{Sn}}\)) on the properties of deposited films. Among the various growth parameters, we found that one sole parameter—Sn flux variation, controlled by adjusting the Sn evaporation temperature of the effusion cell—had a significant impact on the film's properties. Specifically, this adjustment played a crucial role in enhancing the dominance of the FeSn phase, far more effectively than other parameters like substrate temperature or Fe evaporation temperature (\(\mathrm{T_{Fe}}\)). We present a comprehensive study that combines high-resolution X-ray diffraction (HR-XRD), reflection high energy electron diffraction (RHEED), scanning electron microscopy (SEM), transmission electron microscopy (TEM) and vibrating sample magnetometry (VSM) to investigate FeSn thin films grown on GaAs(111) substrates across a range of \(\mathrm{T_{Sn}}\). 

\section{\label{sec:level2}Experimental details}

The films were grown using MBE. Fe and Sn fluxes were generated by a high-temperature effusion cell (HTEZ) and standard effusion cell (WEZ) from MBE Komponenten, respectively.  The films were deposited on 2\textquotedbl\  GaAs (111) substrates.  Before starting the growth, the substrate was heated up to nominally \SI{700}{\celsius} in the growth chamber for 10 minutes to remove oxide layers, residual contaminants and moisture. The substrate is then cooled down to \SI{325}{\celsius} and kept at this temperature during deposition.  A combination of ion-getter pump, cryo pump and LN2 cooling was used to maintain a background pressure of around \(10^{-10}\) Torr during the growth. The deposition time (3 hours) and the substrate rotation speed (5 rpm) were kept constant for all the samples. To investigate the impact of \(\mathrm{T_{Sn}}\), films were deposited at different \(\mathrm{T_{Sn}}\): \SI{950}{\celsius}, \SI{975}{\celsius}, \SI{1000}{\celsius}, \SI{1010}{\celsius}, \SI{1025}{\celsius} and \SI{1035}{\celsius}, while keeping \(\mathrm{T_{Fe}}\) constant at \SI{1350}{\celsius}. A STAIB instrument RH-30 RHEED system was used to monitor the growth process. 

HR-XRD measurements, as presented in figures \ref{fig:GaAs_XRD}, were performed to determine the phase composition and crystalline quality of the films. The diffraction patterns were obtained using a \(\mathrm{Cu- k\alpha}\) (\(\mathrm{\lambda = 1.5406}\) Å) radiation source on a Bruker AXS D8 Discover diffractometer with a half-circle geometry (i.e., \(\mathrm{\omega}\), \(\mathrm{2\theta }\) between \(0^\circ\) and \(180^\circ\)). The crystalline quality and epitaxial alignment of the films were further evaluated through rocking curves obtained using a triple axis analyzer. For each film, the rocking curve was measured for the FeSn(002) peak, with the full width at half maximum (FWHM) providing a measure of the film's mosaicity and epitaxial quality. 

The surface morphology of the films was investigated using Zeiss Ultra 55 FEG-SEM. Transmission electron microscopy (TEM) was used to obtain the cross-section image of the film and scanning transmission electron microscopy - energy dispersive X-ray spectroscopy (STEM-EDS) was used to obtain elemental maps at the film-substrate interface. The data was acquired on a probe and image corrected JEOL ARM200CF, equipped with a Centurio SDD EDS detector. The samples for TEM analysis were prepared by standard focused ion beam (FIB) lift-out technique using a Thermo Scientific Helios G4 UX DualBeam FIB-SEM . The magnetic hysteresis properties of the films were characterized using MicroMag 3900 VSM, with a magnetic field of 1 Tesla applied in the plane of the samples.

\section{Structural characterization}

\begin{figure}[htb!]
    \centering
    \includegraphics[width=0.5\textwidth]{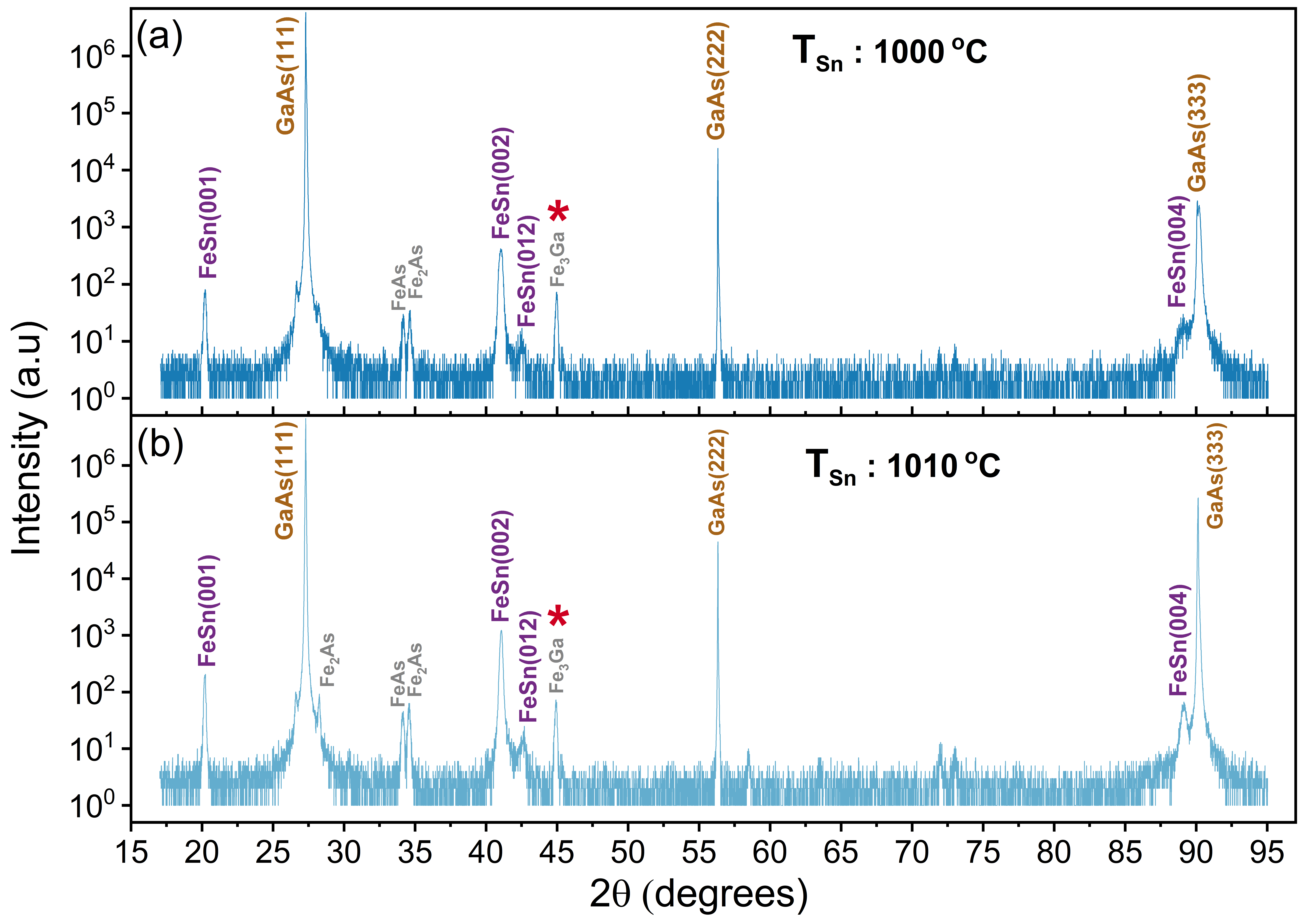}\hfill
     \includegraphics[width=0.5\textwidth]{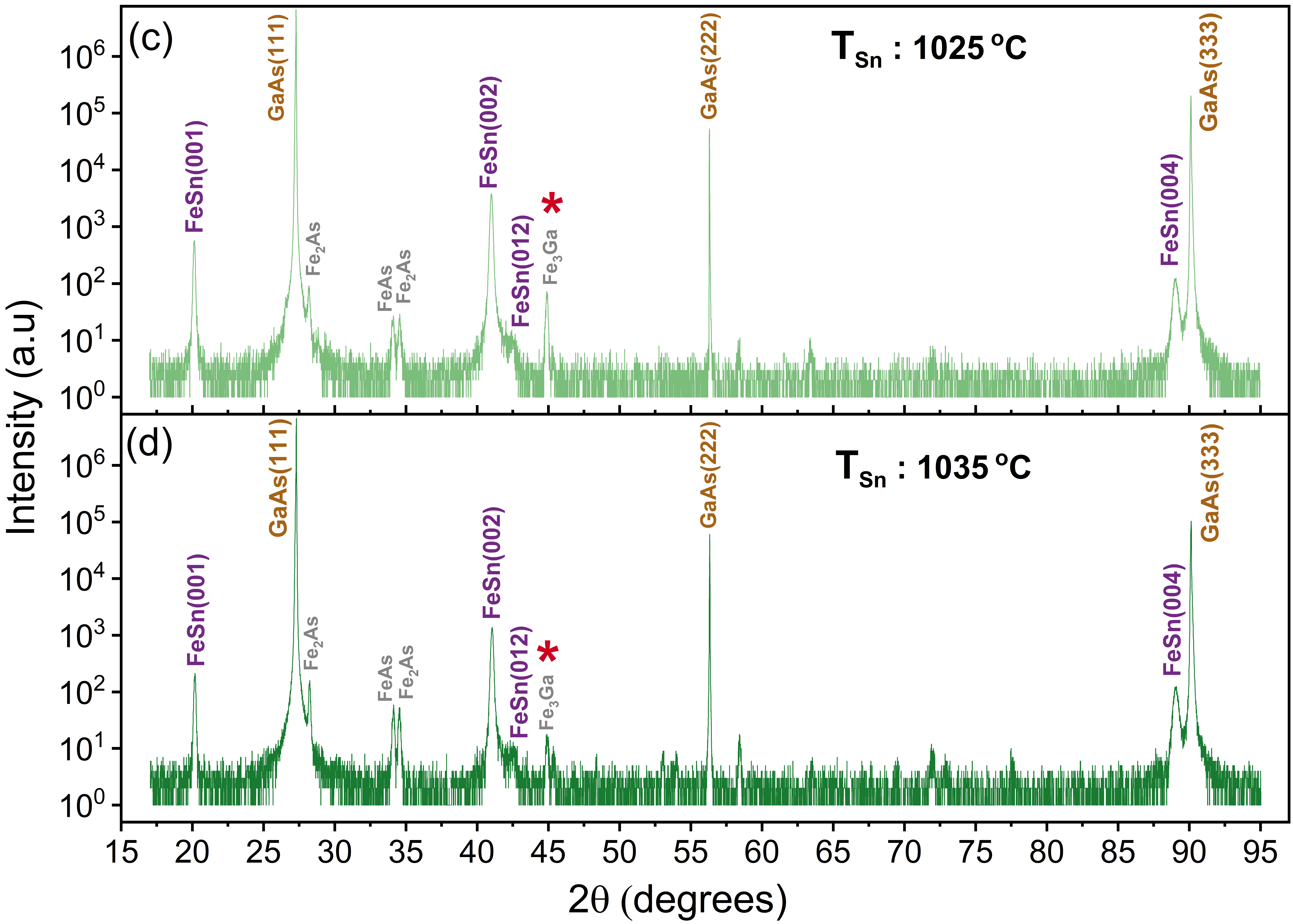}\hfill
    \caption{\textbf{\(\mathrm{2\theta-\omega}\) scans of films deposited on GaAs(111):} \textbf{(a)} At \(\mathrm{T_{Sn}}\) = \SI{1000}{\celsius}, the XRD pattern shows FeSn(001), FeSn(002), and FeSn(004) peaks, confirming the c-axis orientation of the FeSn phase. The additional peaks indicate the presence of secondary phases like \(\mathrm{FeAs}\), \(\mathrm{Fe_2As}\) and  \(\mathrm{Fe_3Ga}\).  As  \(\mathrm{T_{Sn}}\) increases to \textbf{(b)} \SI{1010}{\celsius} and \textbf{(c)} \SI{1025}{\celsius}, there is no significant change in relative peak intensities of FeSn and secondary phases. \textbf{(d)} At \(\mathrm{T_{Sn}}\) = \SI{1035}{\celsius}, the intensity of the \(\mathrm{Fe_3Ga}\) peak (marked with \textcolor{red}{*}), an undesirable ferromagnetic phase, significantly reduces.}
    \label{fig:GaAs_XRD}
\end{figure}

Figure \ref{fig:GaAs_XRD} shows the HR-XRD \(\mathrm{2\theta - \omega}\)  scans of FeSn thin films grown on GaAs(111) substrates as a function of \(\mathrm{T_{Sn}}\). The presence of FeSn(001), FeSn(002), and FeSn(004) peaks confirms the c-axis orientation of  FeSn. A low-intensity FeSn(012) peak is also observed. The average out-of-plane lattice constant is 4.4 Å. The FeSn(002) peak appears at \( 2\theta\approx40.9^\circ\) and the relative intensities between the FeSn peaks are consistent with the previously reported values \cite{inoue2019molecular,hong2020molecular,khadka2020high,bhattarai2023magnetotransport}. In addition to FeSn peaks, peaks corresponding to secondary phases \(\mathrm{FeAs}\), \(\mathrm{Fe_2As}\) and \(\mathrm{Fe_3Ga}\), are observed.  

At lower \(\mathrm{T_{Sn}}\)  values, the XRD spectra show clear peaks corresponding to these secondary phases. As \(\mathrm{T_{Sn}}\) increases, while the intensity of these secondary phase peaks does not decrease drastically,  a significant reduction in the \(\mathrm{Fe_3Ga}\) peak (marked with \textcolor{red}{*}) intensity is observed at \(\mathrm{T_{Sn}}\) = \SI{1035}{\celsius}. \(\mathrm{Fe_3Ga}\) is ferromagnetic, which is undesirable in conjunction with the intended antiferromagnetic properties of FeSn. Additionally, the relative intensity of the FeSn(012) peak decreases with increasing \(\mathrm{T_{Sn}}\). The reduction of both \(\mathrm{Fe_3Ga}\) and FeSn(012) peaks, along with a sharper and slightly more intense FeSn(002) peak, suggests that higher \(\mathrm{T_{Sn}}\) not only improves the crystalline quality of the FeSn phase but also reduces the formation of unwanted ferromagnetic secondary phases. This transition is further supported by the magnetic hysteresis measurements, discussed in section \ref{VSM}. 

\begin{figure}[htb!]
    \includegraphics[width=0.5\textwidth]{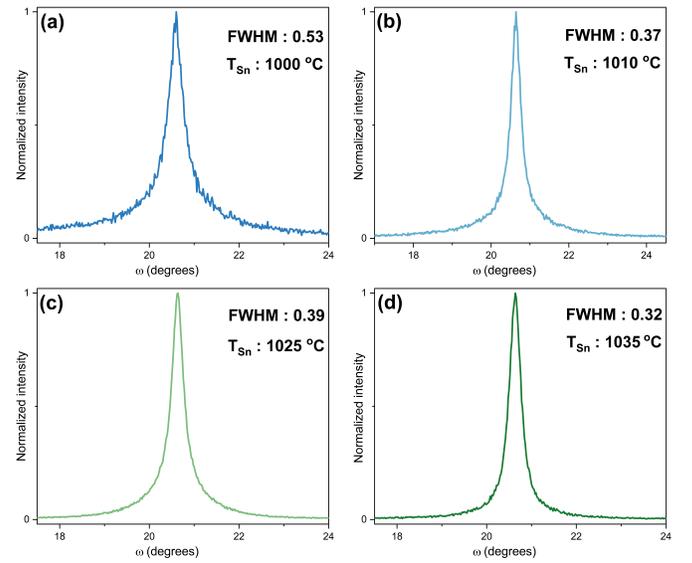}\hfill
    \caption{\textbf{Rocking curve analysis of films deposited on GaAs(111): }\textbf{(a)} At \(\mathrm{T_{Sn}}\) = \SI{1000}{\celsius}, the FWHM of \(0.53^\circ\) indicates a reasonably good degree of crystal orientation, though it suggests the presence of some lattice strain or defects. At \textbf{(b)} \(\mathrm{T_{Sn}}\) = \SI{1010}{\celsius} and \textbf{(c)} \(\mathrm{T_{Sn}}\) = \SI{1025}{\celsius} the FWHM narrows to \(0.37^\circ\) and \(0.39^\circ\) respectively, pointing to a slight improvement in the uniformity of crystal orientation and a potential relaxation in strain. \textbf{(d)} At \(\mathrm{T_{Sn}}\) = \SI{1035}{\celsius}, the FWHM further decreases to \(0.32^\circ\), indicating an improvement in the crystal quality of the film. Although the differences in FWHM are subtle, they suggest that the film maintains good epitaxial quality across the range of \(\mathrm{T_{Sn}}\) values.}
    \label{fig:GaAs_RockingCurve}
\end{figure}

The rocking curves of the FeSn(002) peak in figure \ref{fig:GaAs_RockingCurve} provide further insight into the role of  \(\mathrm{T_{Sn}}\) in refining the epitaxial quality of the FeSn films. At \(\mathrm{T_{Sn}}\) = \SI{1000}{\celsius}, the FWHM of FeSn(002) peak is \(0.53^\circ\), indicating a reasonably good degree of crystal quality. However, it suggests the presence of some lattice strain or defects. As \(\mathrm{T_{Sn}}\) increases to \SI{1010}{\celsius} and \SI{1025}{\celsius}, the FWHM narrows slightly to \(0.37^\circ\) and \(0.39^\circ\), respectively. This slight narrowing suggests a modest improvement in the uniformity of crystal orientations and a potential relaxation of strain within the films. 

At  \(\mathrm{T_{Sn}}\) = \SI{1035}{\celsius}, the FWHM decreases to \(0.32^\circ\). While this represents the smallest FWHM value observed, indicating a slightly enhanced epitaxial alignment, it is important to note that the differences between the FWHM of the last three samples are relatively small. The improvement in FWHM with increasing  \(\mathrm{T_{Sn}}\) is consistent but not drastic. This suggests that the epitaxial quality of the films is reasonably good even at low \(\mathrm{T_{Sn}}\) values and that the primary trend is a subtle refinement of crystal orientation as \(\mathrm{T_{Sn}}\) increases. XRD data for the films grown at \SI{950}{\celsius}  and \SI{975}{\celsius} are not shown in the structural characterization since they predominantly show the presence of only secondary phases \(\mathrm{Fe_{}As}\), \(\mathrm{Fe_{2}As}\) and \(\mathrm{Fe_{3}Ga}\) and not FeSn. 

\section{Surface morphology and elemental mapping}

\begin{figure}[htb!]
    \includegraphics[width=0.5\textwidth]{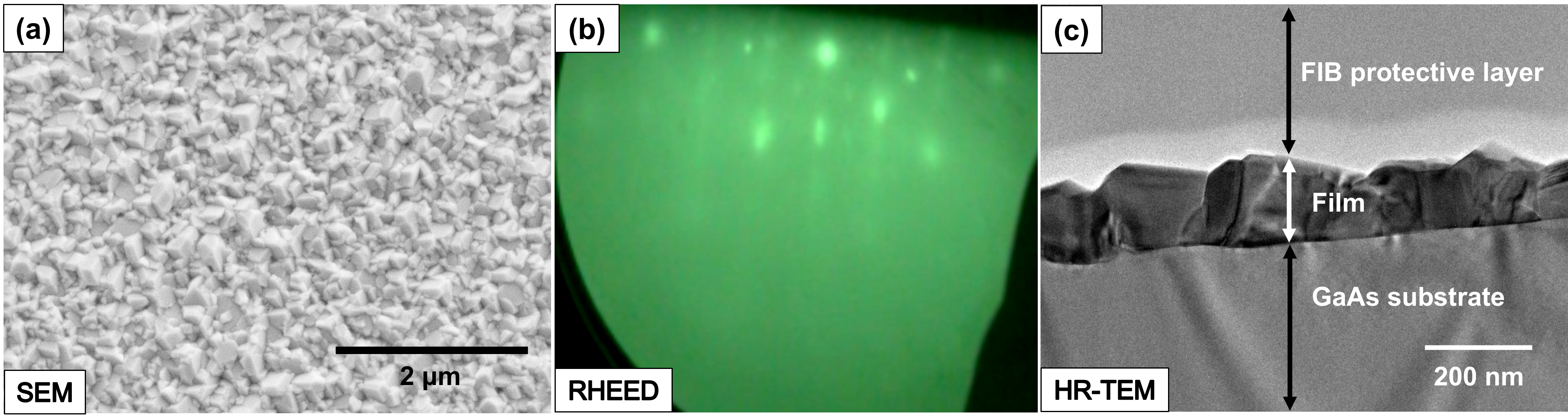}\hfill
    \includegraphics[width=0.5\textwidth]{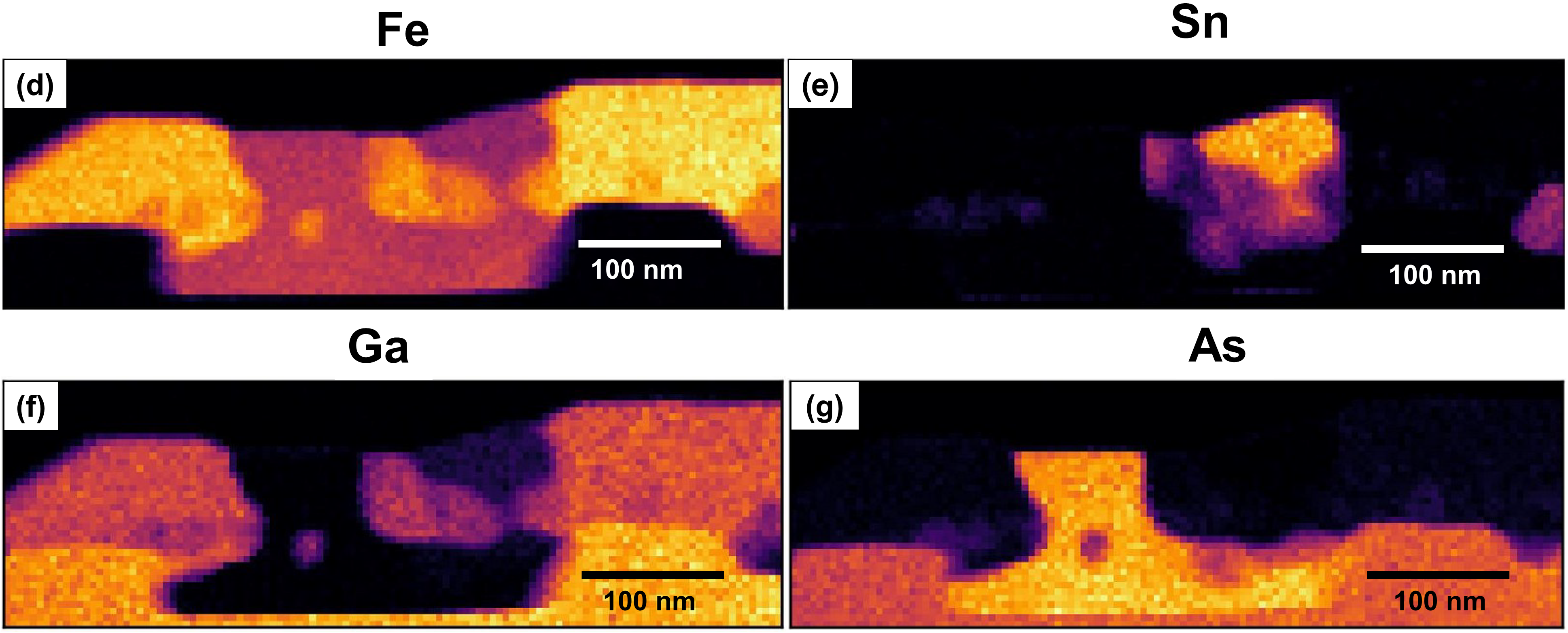}\hfill
    \caption{\textbf{Surface morphology, atomic-scale elemental mapping of FeSn/GaAs interface:} \textbf{(a)} SEM image of the film deposited at \(\mathrm{T_{Sn}}\) = \SI{1035}{\celsius}, revealing a rough textured surface. \textbf{(b) } The transmission type RHEED pattern recorded during the growth validates the presence of 3D islands on the surface. \textbf{(c)} The cross-section TEM image of the same film shows the thickness is approximately 200 nm.\textbf{ (d)} The EDS elemental map of Fe shows a consistent presence of Fe across the film, indicating the formation of Fe-based compounds.  \textbf{(e)} The Sn map shows the presence of Sn-rich regions that overlap with Fe, suggesting the formation of FeSn islands. \textbf{(f,g)} The elemental maps of Ga and As overlap with Fe but not with Sn. This indicates inter-diffusion of Fe, Ga and As.}
    \label{fig:GaAs_SEM_TEM}
\end{figure}

Among the six samples grown at different \(\mathrm{T_{Sn}}\), the film grown at \(\mathrm{T_{Sn}}\) = \SI{1035}{\celsius} was chosen for further investigation due to the favourable results seen in the XRD analysis. The XRD data of this film shows a significant suppression of the peak corresponding to ferromagnetic \(\mathrm{Fe_3Ga}\) and the FWHM of the FeSn(002) peak is the smallest compared to the other three films. 

The surface morphology and elemental composition of the deposited films are critical for understanding the growth mechanism, phase formation, and chemical diffusion at the film-substrate interface. Surface roughness provides insights into the growth mode, such as whether the film grows layer-by-layer or forms 3D islands. Elemental mapping is essential for understanding the inter-diffusion of Fe, Ga and As, which contribute to the formation of secondary phases. We used SEM to investigate the surface morphology and STEM-EDS was used for elemental mapping of the film's cross-section. 

In figure \ref{fig:GaAs_SEM_TEM}(a), the SEM image gives an overview of the surface morphology of the film grown at  \(\mathrm{T_{Sn}}\) = \SI{1035}{\celsius}. The granularity and non-uniformity of the surface suggest that the growth is not layer-by-layer, but rather three-dimensional where islands form and merge as the growth proceeds. The RHEED pattern (figure \ref{fig:GaAs_SEM_TEM}(b)) observed during the growth is consistent with the surface morphology results from SEM. The observed transmission-type diffraction pattern corresponds to a rough surface with three-dimensional islands. 

Figure \ref{fig:GaAs_SEM_TEM}(c) is the cross-sectional TEM image of the same film. The average thickness of the film is around 200 nm. The EDS elemental maps in figure \ref{fig:GaAs_SEM_TEM}(d)-(g) represent cross-sectional views of the same region as the TEM image in figure \ref{fig:GaAs_SEM_TEM}(c). In the elemental maps, bright regions indicate a high concentration of the corresponding element, while dark regions indicate a low concentration. It is important to note that these maps may not fully represent the exact elemental distribution across the entire film. These localized elemental maps provide a snapshot of the elemental distribution in a specific cross-section and a general understanding of the interdiffusion and possible compound formation based on elemental overlap. 

The elemental maps, illustrated in figure \ref{fig:GaAs_SEM_TEM}(d), (f) and (g), show an overlap between Fe and the diffused elements Ga and As from the substrate. The presence of Fe across the film, as seen in the EDS map in figure \ref{fig:GaAs_SEM_TEM}(d), indicates the formation of Fe-based compounds. It also shows the diffusion of Fe into the substrate. This aligns with the XRD data, which shows the presence of FeSn and secondary phases, such as  \(\mathrm{Fe_{}As}\), \(\mathrm{Fe_{2}As}\) and \(\mathrm{Fe_{3}Ga}\). Conversely, the Sn map depicted in figure \ref{fig:GaAs_SEM_TEM} (e) appears less uniformly distributed, showing regions with higher concentrations of Sn. These Sn-rich regions also overlap with Fe, which suggests the formation of FeSn islands. The Sn-rich regions appear smaller compared to other elements in this specific cross-sectional view. As mentioned earlier, this mapping does not fully represent the Sn distribution across the entire film and these elemental maps should not be used for quantitative analysis.

\section{Magnetic properties}
\label{VSM}

\begin{figure}[htb!]
    \includegraphics[width=0.5\textwidth]{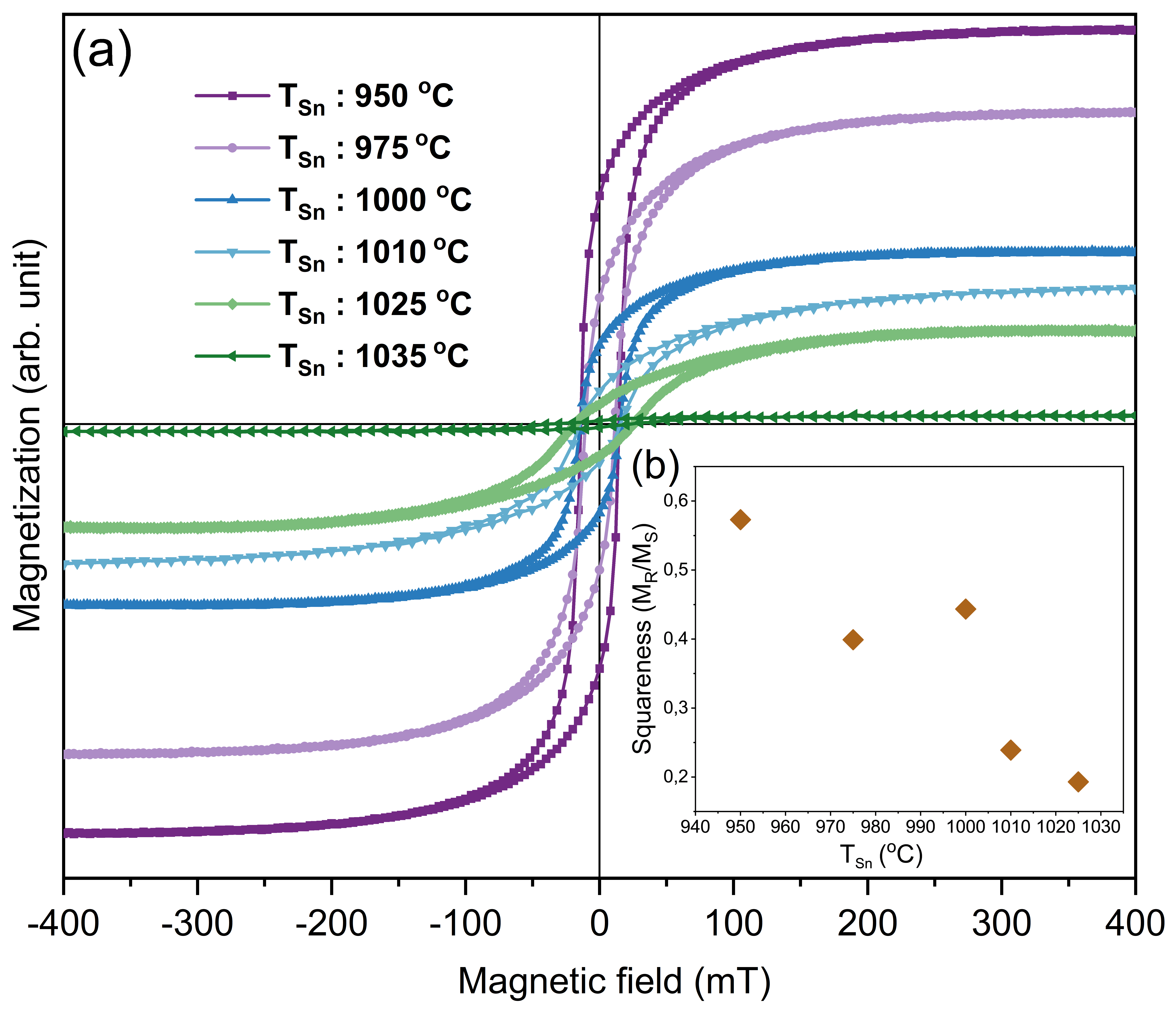}\hfill
    \caption{\textbf{Sn evaporation temperature dependent magnetic properties: } \textbf{(a)} Hysteresis loops for films grown at varying  \(\mathrm{T_{Sn}}\) ranging from \(\mathrm{T_{Sn}}\) = \SI{950}{\celsius} to \(\mathrm{T_{Sn}}\) = \SI{1035}{\celsius}. The loops exhibit a decrease in squareness (\(\mathrm{M_r/M_s}\)) with increasing \(\mathrm{T_{Sn}}\). \textbf{(b) }The inset graph quantitatively depicts the decline in \(\mathrm{Mr/M_s}\) as a function of \(\mathrm{T_{Sn}}\). }
    \label{fig:VSM}
\end{figure}

To complement the structural and compositional analysis of the films, it is essential to evaluate their magnetic properties. Variations in \(\mathrm{T_{Sn}}\) were shown to influence the crystallinity of the films, particularly concerning the suppression of the ferromagnetic phase \(\mathrm{Fe_{3}Ga}\). Magnetic characterization was performed by VSM to correlate these structural changes with the films' magnetic behaviour. The primary aim of this characterization is to examine how the increasing influence of the antiferromagnetic FeSn phase at higher \(\mathrm{T_{Sn}}\) affects the magnetic response. These measurements provide insight into the extent to which \(\mathrm{T_{Sn}}\) optimizes antiferromagnetic ordering by minimizing the presence of ferromagnetic impurities. 

The series of hysteresis loops, shown in figure \ref{fig:VSM}(a), illustrates how \(\mathrm{T_{Sn}}\) influences the magnetic properties of the films. As \(\mathrm{T_{Sn}}\) increases from \SI{950}{\celsius} to \SI{1025}{\celsius}, the squareness (\(\mathrm{M_r/M_s}\)) of the hysteresis loops decreases from 0.57 to 0.19, which is consistent with the expected antiferromagnetic nature of the film. This trend can be attributed to the increasing influence of the antiferromagnetic FeSn over ferromagnetic \(\mathrm{Fe_{3}Ga}\). The inset graph in figure \ref{fig:VSM}(b) highlights the variation of \(\mathrm{M_r/M_s}\) with \(\mathrm{T_{Sn}}\), showing a downward trend as \(\mathrm{T_{Sn}}\) rises. 

For the film grown at \(\mathrm{T_{Sn}}\) = \SI{1035}{\celsius}, the hysteresis loop becomes almost a straight line, showing no noticeable coercivity or remanence. Due to this behaviour, the squareness (\(\mathrm{M_r/M_s}\)) of the loop cannot be calculated, as there is no measurable remanent magnetization (\(\mathrm{M_r}\)). The nearly linear response to the applied magnetic field could result from a paramagnetic or non-magnetic sample. However, it can be more likely attributed to the antiferromagnetic nature of the film, given that the observation aligns well with the XRD data, which shows suppression of the ferromagnetic \(\mathrm{Fe_3Ga}\) phase at \(\mathrm{T_{Sn}}\) = \SI{1035}{\celsius}. 

\section{Conclusion}

In this study, we investigated the epitaxial growth of FeSn thin films on GaAs(111) substrates using MBE, with a focus on the influence of  \(\mathrm{T_{Sn}}\) on the film's structural and magnetic properties. The challenges posed by the formation of secondary phases such as \(\mathrm{FeAs}\), \(\mathrm{Fe_2As}\), and \(\mathrm{Fe_3Ga}\) due to the inter-diffusion of Fe, Ga and As, are significant. However, our results show that careful optimization of \(\mathrm{T_{Sn}}\) can substantially enhance the crystallinity and phase purity of the FeSn films.

Our comprehensive characterization, including HR-XRD, SEM, TEM, and VSM, demonstrates that increasing \(\mathrm{T_{Sn}}\) leads to a reduction in the unwanted ferromagnetic \(\mathrm{Fe_3Ga}\) phase and a corresponding increase in the influence of the antiferromagnetic FeSn phase. While perfect single-phase FeSn films are challenging to achieve on GaAs substrates due to the diffusion issues, our findings highlight the critical role of \(\mathrm{T_{Sn}}\) in steering the growth towards the desired FeSn phase.

These insights into the relationship between growth conditions and film properties are vital for advancing the fabrication of FeSn thin films. Future work could explore further optimization of growth parameters and alternative substrate treatments to mitigate diffusion effects and achieve even higher phase purity.

\begin{acknowledgments}
This work was supported by the Research Council of Norway: through its Centres of Excellence funding scheme, QuSpin (262633), the Norwegian Center for Transmission Electron Microscopy, NORTEM (197405), the Norwegian Micro- and Nano-Fabrication Facility, NorFab (295864) and In-situ Correlated Nanoscale Imaging of Magnetic Fields in Functional Materials, InCoMa (315475). J.H. and D.M. acknowledge funding from the European Research Council (ERC) under the European Union’s Horizon 2020 Research and Innovation Program (Grant Agreement No.863691).
\end{acknowledgments}

\section*{Author Declarations}
\subsection*{Conflict of interest}
The authors declare no conflicting interests.

\section*{Data Availability Statement}
The data that support the findings of this study are available from the corresponding author upon reasonable request.

\providecommand{\noopsort}[1]{}\providecommand{\singleletter}[1]{#1}%
%

\end{document}